\documentclass[journal=nalefd,manuscript=article]{achemso}
\author{Aurélien Masseboeuf}
\affiliation[CEA,Institut Nanosciences et Cryogénie - SP2M, 17 rue des Martyrs, 38054 Grenoble Cedex 09, France]{CEA,INAC, Grenoble (FR)}
\altaffiliation{New ad. : CEMES-CNRS, Toulouse, France}
\email{aurelien.masseboeuf@cemes.fr}
\author{Alain Marty}
\affiliation[CEA,Institut Nanosciences et Cryogénie - SP2M, 17 rue des Martyrs, 38054 Grenoble Cedex 09, France]{CEA,INAC, Grenoble (FR)}
\author{Pascale Bayle-Guillemaud}
\affiliation[CEA,Institut Nanosciences et Cryogénie - SP2M, 17 rue des Martyrs, 38054 Grenoble Cedex 09, France]{CEA,INAC, Grenoble (FR)}
\author{Christophe Gatel}
\affiliation[CEMES-CNRS, BP 94347, 29 rue Jeanne Marvig, 31055 Toulouse Cedex, France]{CEMES-CNRS, Toulouse (FR)}
\author{Etienne Snoeck}
\affiliation[CEMES-CNRS, BP 94347, 29 rue Jeanne Marvig, 31055 Toulouse Cedex, France]{CEMES-CNRS, Toulouse (FR)}

\title[Quantitative magnetic observation of PMA alloys at the nanoscale]
{Quantitative observation of magnetic flux distribution in new magnetic films for future high density recording media.}

\begin{document}
\begin{figure}[!h]
\includegraphics[width = 8cm]{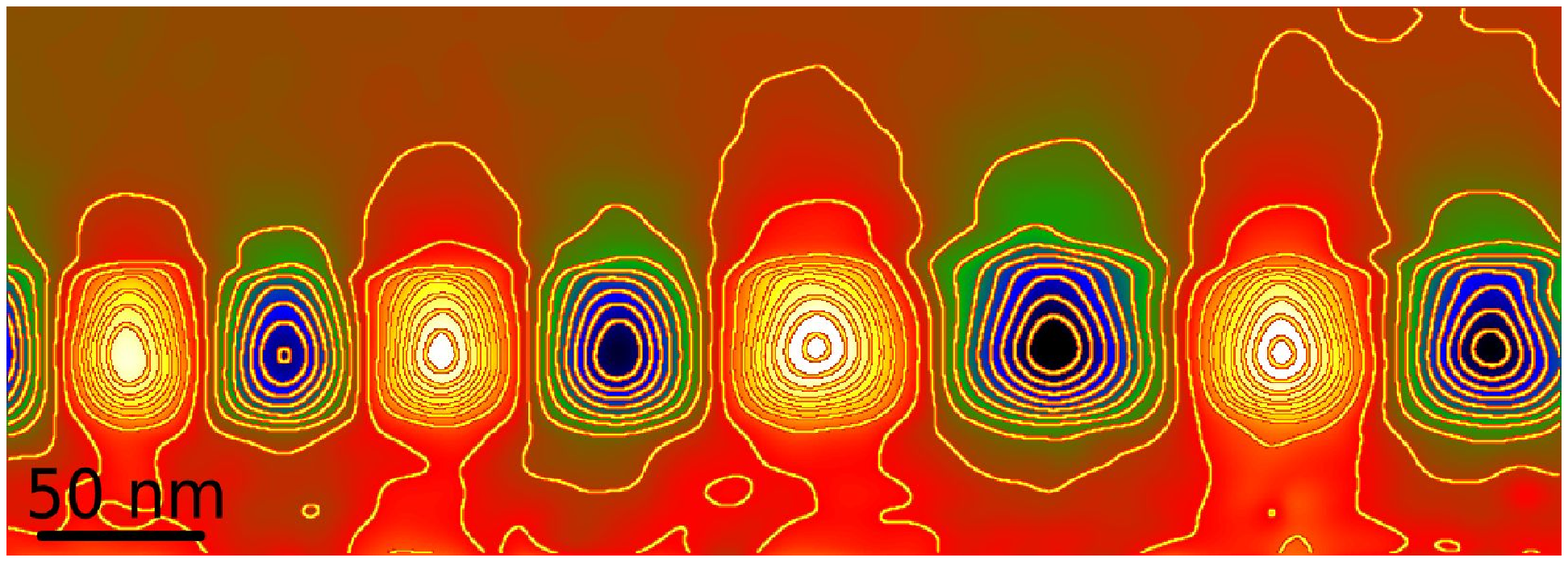}
\label{FigTOC}
\end{figure}

\begin{abstract}
Off-axis electron holography was used to observe and quantify the magnetic microstructure of a perpendicular magnetic anisotropic (PMA) recording media. Thin foils of PMA materials exhibit an interesting Up and Down domain configuration. These domains are found to be very stable and were observed at the same time with their stray field, closing magnetic flux in the vacuum. The magnetic moment can thus be determined locally in a volume as small as few tens of nm$^3$. 
\end{abstract}



From the first desktop (5.25 inch diameter) hard disk drive (HDD) in the early 1980's, with 5 MegaBytes storage capacity, to the 1 TeraByte hard drive distributed now by Hitachi, fundamental improvments have been achieved in HDD technology. The main evolutionnary step in the data storage history happened certainly before the first HDD-integrated Personal desktop Computer (Seagate in 1980) when the perpendicular recording idea was first introduced in 1977.\cite{iwasaki_analysis_1977} Now, all improvements in new memory capacity are expected to be reached thanks to perpendicular recording. While important studies are published on new materials design, \cite{kryder_high-density_2005,Richter2007} studies on magnetic interaction between the recording head  and the data bits are mostely concerned about the tip field leakage characterization \cite{Signoretti2004,kim2008}. Here we show that it is also possible to obtain nanometric and quantitative magnetic informations of stray field and magnetic induction at the same time on perpendicular magnetic anisotropic (PMA) materials.\\

Due to the strong stray fields perpendicular to their surface, PMA materials have been extensively studied by Magnetic Force Microscopy (MFM),\cite{gehanno_magnetic_1997,attane_domain_2001} micro-magnetic simulations,\cite{toussaint_new_2002} and other magnetic characterization techniques.\cite{durr1999,beutier_soft_2004} However these techniques were not suitable to study the inner magnetic configuration of materials and inner magnetization at the same time. Thus, formation of stray field distribution with respect inner magnetic parameters of the material have not been yet confirmed by experimental results\cite{Hubert1998}. This could be of great importance to evaluate the ability of a reading head to flip a bit.\\

Electron Holography (EH) in a Transmission Electron Microscope (TEM) is a powerful technique which enables observation of electrostatic and magnetic fields at the nanoscale by a electron wave phase retrieval process.\cite{gabor_new_1948} Through the so-called Aharonov-Bohm effect,\cite{aharonov_significance_1959} it is known that an electron wave is sensitive to the electric and magnetic potential. As a consequence, it can be used to investigate magnetic properties of materials. Electron holography has thus been used to study many magnetic materials, for example, the analysis of stray fields around a MFM tip,\cite{streblechenko_quantitative_1996} or the study of the magnetic configuration of magnetic nanoparticles,\cite{Hytch2003}  magnetic films,\cite{mccartney_off-axis_1997} magnetic tunnel junctions,\cite{Snoeck2008} or even magnetite core of magnetotactic bacteria.\cite{dunin-borkowski_magnetic_1998} The phase shift of the exit electron wave, $\Delta \phi$, travelling along the \emph{z} direction across the sample, which interacted with the electromagnetic field (electrostatic and magnetic potentials from the sample) can be expresses as :
\begin{equation}
\label{Eq1}
			\Delta \phi\left( x,y\right)  = C_{E} \int V_{int}(x,y,z)  dz - \frac{e}{\hbar}\int A_{z}(x,y,z)  dz
\end{equation}
where $V(x)_{int}$ represents the electrostatic contribution to the phase shift (in the case of a material it is mainly its Mean Inner Potential or MIP), and  $A_{z}$ is the \emph{z}-component of the magnetic vector potential describing the magnetic induction distribution in a plane (for a given z) pendicular to the optic axis. $A_z$ is related to the magnetic induction $\vec{B}$ by means of the Maxwell's equation : $\vec{B} = (B_x,B_y) = (\nabla_y A_z,-\nabla_x A_z)$. $C_{E}$ is an electron energy related constant.\\
The keypoint is the separation of the magnetic and electrostatic contributions in the reconstructed phase shift and is extensively discussed elsewhere.\cite{dunin-borkowski_off-axis_2004}

We have used for our purpose a method consisting of recording two holograms before ($\Delta \phi_{+}$) and after ($\Delta \phi_{-}$) removing and inverting the sample. The electrostatic contribution to the phase shift remains similar in the two holograms while the magnetic contribution changes in sign. The magnetic contribution ($\Delta \phi_{magn.}$) can thus be obtained by evaluating half of the difference of the two phase images calculated from the two holograms. The MIP contribution ($\Delta \phi_{MIP}$) is then half of the sum. To account for the sample reversal, it is necessary to reverse one of the two phase images and align them.
\begin{equation}
\begin{array}{rcl}
   \label{Eq2} 
	\Delta \phi_{magn.}  = & \frac{1}{2}*\left[ \Delta \phi_{+} - \Delta \phi_{-}\right]  & =  \frac{e}{\hbar} \left[A_{z}\Delta t(x,y)\right]  \\
   	\Delta \phi_{M.I.P.} = & \frac{1}{2}*\left[ \Delta \phi_{+} +\Delta  \phi_{-}\right]  & = C_{E} V(x,y)_{int} \Delta t(x,y) 
\end{array}
\end{equation}
EH has been used to study the magnetic configuration and measure the remanent magnetization of a FePd$ _{L1_{0}}$/FePd$ _{disord.}$ stack, grown on MgO(001), which exhibits a strong PMA. The tetragonal axis of the chemically ordered FePd$_{L1_{0}}$ crystalline structure lies along the growth direction corresponding to an alternate stacking of pure Iron and Palladium planes. This chemical anisotropy along the \textit{z}-axis induces an easy magnetic axis in the same direction which gives rise to the up and down magnetic domain configuration.\cite{gehanno_magnetic_1997} The main purpose of the second "soft" layer, having a vanishing anisotropy, is usually described as the main factor for increasing the recording efficiency.\cite{khizroev_perpendicular_2004}\\
The expected magnetic configuration of the FePd$ _{L1_{0}}$/FePd$ _{disord.}$ bilayer is presented in Figure1-A. Domains in the FePd$ _{L1_{0}}$ layer are separated by Bloch walls where the magnetization lies in the plane of the foil, surrounded by a N\'eel Cap in which the magnetization runs around the Bloch wall axis. The bottom FePd$ _{disord.}$ layer gives rise to in-plane components allowing a flux closure within the bilayer, and enables the domains to be aligned in a parallel stripes configuration. This magnetic configuration is confirmed by studying the external stray field by MFM experiment as shown in Figure 1-B.\\
\begin{figure}
  \includegraphics[width = 0.7\columnwidth]{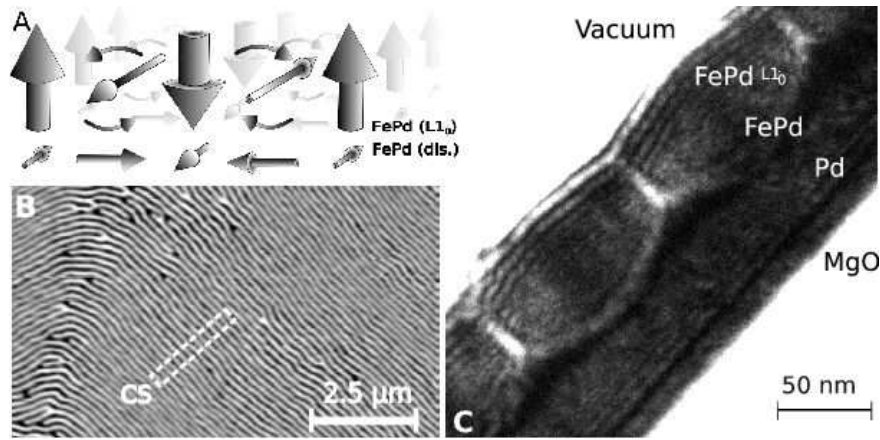}
  \caption{\textbf{A.} Magnetic configuration expected for the foil. \textbf{B.} MFM view of the sample in its stripe configuration. Black contrast is down domains, bright contrast is up ones. The dashed area shows the geometry used for TEM sample preparation, across magnetic orientation. \textbf{C.} Fresnel view of the sample showing black and white contrast where the beam are overlaping.}
  \label{Fig1}
\end{figure}

Our purpose is to analyse in more detail the inner magnetic configuration with higher resolution. Figure 1-C shows a Fresnel TEM micrograph\cite{chapman_investigation_1984} of the sample. The thin foil used for TEM experiments has been prepared in cross-section in order to observe the magnetic structure by the side instead of the top in the MFM geometry. In this micrograph, domains are clearly defined, separated from one another by a bright or dark line localised at the position of domain walls. The domain periodicity is 100 nm which fully agree with MFM measurement. EH experiments were carried out on the same area of the PMA magnetic film.\\
\begin{figure}
\includegraphics[width = 0.8\textwidth]{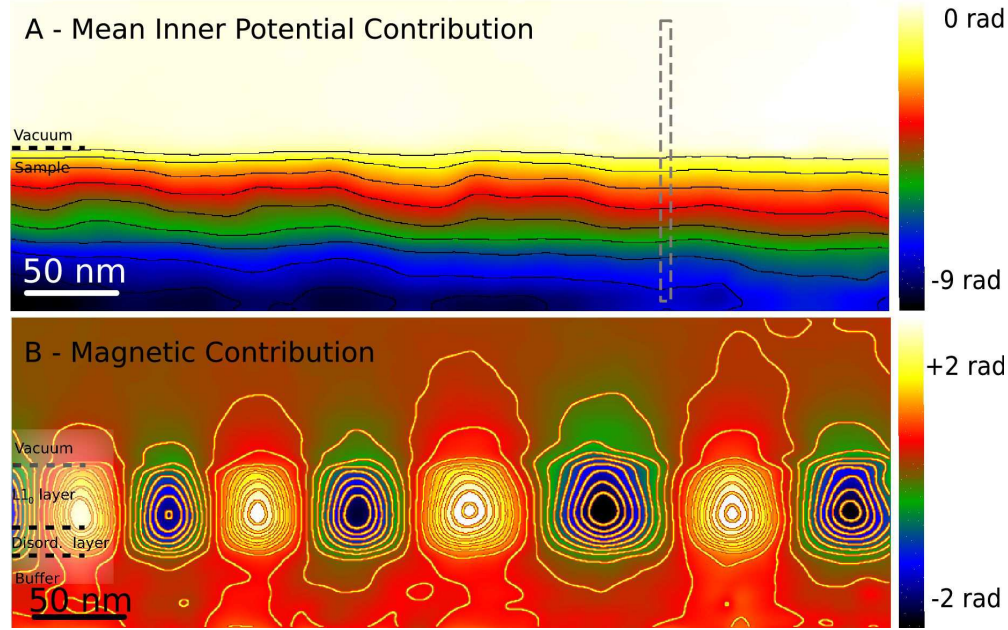}
\caption{Mean Inner Potential (\textbf{A}) and magnetic (\textbf{B}) contribution to the phase of the electron wave. The color scale used here is a temperature scale described near each picture. The contour lines are for equi-phase lines and represent 1 radian for MIP contribution and  1/4 radian for magnetic contribution.}
\label{Fig2}
\end{figure}

Figure 2-A shows the deduced MIP contribution to the phase shift, and the magnetic contribution is shown in Figure 2-B. The iso-phase contours displayed on both phase images directly relate to thickness variations (in the MIP phase image) and to magnetic flux (in the magnetic one). The variation of the MIP contribution, exhibits that the TEM sample increases uniformly in thickness while magnetic contribution highlights vortices corresponding to the Bloch walls. Between these vortices are areas where the magnetic flux is parallel or anti-parallel to the growth direction. These correspond to the "up" and "down" magnetic domains. Stray fields close the flux in the vacuum and inside the stack. However, the vortices appear to be flatter at the bottom (close to the FePd disordered layer) than at the top (close to the vaccum). This asymmetrical shape of the vortices is due to the disordered FePd layer which forces the magnetic induction to lie within the foil plane. 
From equation \ref{Eq2}, quantitative values of magnetic induction can be extracted provided that the MIP or the thickness of the different layers are known:
 \begin{eqnarray}
\begin{array}{rcl}
   \label{Eq3} 
	\Delta t(x) &=& \frac{\Delta \phi_{M.I.P.}}{C_{E} V(x)_{int}}\\
	B_{\perp} &=& \frac{\hbar}{e\cdot\Delta t(x)} \nabla [ \Delta \phi_{magn.} ] 
\end{array}
\end{eqnarray}

\begin{figure}
\includegraphics[width = 0.5\columnwidth]{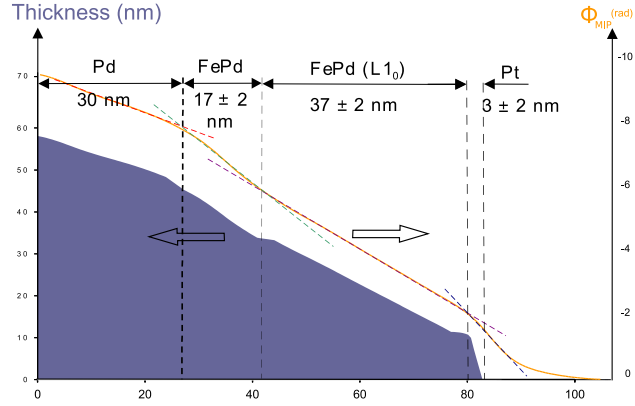}
\caption{Mean Inner Potential contribution to the phase analysis. This is a plot profile along the dashed line in Figure 2-A. Right (yellow) label and solid line is the phase profile used to deduce the different layers in the foil. Dashed lines present linear interpolation variations for each different layers. Thickness profile is presented in the colored (blue) area and is labeled on the left.}
\label{Fig3}
\end{figure}

Figure 3 shows in yellow the experimental profile of the electrostatic contribution to the phase extracted along the dashed line in Figure 2. To deduce the thickness profile (in blue), we have first calculated the mean inner potential values for the Pd, FePd$ _{L1_{0}}$ and FePd$ _{disord.}$ layers.\\
According to Equation 3, this thickness profile is used to calculate the $B_x$ and $B_y$ inside the layer (Figure 4, A and B). Neglecting the demagnetizing field within the material, the measured magnetic induction is directly related to the magnetization in the material. The value of the magnetic induction modulus in the FePd$ _{L1_{0}}$ region (\emph{i.e.} inside the domains) gives rise to a magnetic induction of $1.3 \pm 0.1$ T while same measurements performed on the FePd$ _{disord.}$ area (under domain walls) gives an aeeeveraged values of $1.2 \pm 0.1$ T.  Values measured for the $\mu_{0}M_{s}$ of the FePd$ _{L1_{0}}$ layer are the same as those expected for FePd.\cite{gehanno_magnetic_1997} The variations observed in the different domains come from a variation in the evaluation of the local thickness, due to small deformation of the crystal, or amorphization during ion milling.
The difference found between the two layers is negligible. It should comes from the smaller area of magnetization purely perpendicular to the electron path in the soft layer under the domain wall. This implies a variation of the magnetization direction due to the presence of the domain wall and the in-plane magnetization under the domains. The measured value is then no longer a pure magnetic moment but a projection of it.\\
\begin{figure}
\includegraphics[width = 0.5\columnwidth]{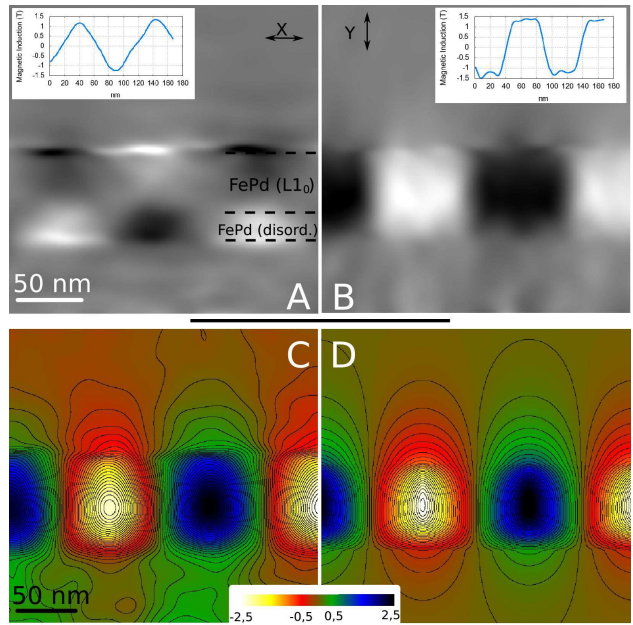}
\caption{\textbf{A} and \textbf{B} are the X and  Y components (left and right respectively) of the magnetic induction deduced from phase gradients. Profile for quantitative interpration is displayed on each figure. \textbf{C} is a zoomed view of Fig. 2-B compared with a micro-magnetic simulation in \textbf{D}. In both images, iso-phase lines represent $0.08 \; rad$ and the color scale used is described.}
\label{Fig4}
\end{figure}
More accurate measurements of the FePd magnetic properties can be done performing micromagnetic simulations. We have used a code based on LLG temporal integration, GL\_FFT,\cite{toussaint_new_2002} to simulate the magnetic flux  (i.e. the iso-phase contours) observed by EH. Calculated magnetic phase shift (using bulk values\cite{gehanno_magnetic_1997}) and experimental data are compared in Figure 4, C and D.\\
It is seen that the Bloch walls are much wider in the experimental data which could be explain by a slight decrease of the thickness due to the thinning process (see also Supporting Informations). The magnetic flux can be quantitatively measured, both in and outside the sample. The stray field can then be related to the magnetization moment in the domains. This is potentially of great interest for the design of a reader of this kind of material.\\

Electron holography was used to highlight the magnetic structure of FePd$ _{L1_{0}}$-FePd$ _{disord.}$ magnetic bilayer exhibiting PMA, with a resolution closed to the nanometer and an accurate measurement of the local magnetic induction. The magnetic configuration was then successfully compared to micromagnetic calculations. Moreover, the quantitative informations given by the technique can be directly related to the stray field of the materials, which are the bit information for reading heads in hard drives.

\begin{acknowledgement}
Thanks are due to Dr. P. Cherns for critics and usefull discussions about this manuscript.
\end{acknowledgement}

\begin{suppinfo}
The sample was grown on MgO (001) substrate by Molecular Beam Epitaxy (MBE) according to the following sequence : Cr (2.5 nm) in order to initiate the epitaxial growth, Pd (48 nm), FePd (15 nm) co-deposited at room temperature, FePd (37 nm) co-deposited at 450$^{\circ}C$ and a 1.5 nm caping of Pt was added to avoid oxydation.\\
The sample has been prepared for electron microscopy using mechanical polishing and ion milling. The layer is thus exhibiting a double wedge geometry along the observation plane. The microscope used for the holography experiments is a FEI Tecnai F20 field-emission-gun TEM fitted with a Cs corrector (CEOS). A FEI Titan FEG TEM fitted with a dedicated Lorentz lens was used for Fresnel imaging. A Gatan Imaging Filter was also used for zero loss filtering for the Fresnel images.\\
Holograms are recorded using off-axis electron holography with a rotatable biprism located in the SA aperture. The biprism is aligned along the foil direction \emph{x}. The fringe spacing is 1.8 nm, the fringe contrast is 12 \%. For calculating the phase image we perform a Fourier transform of the hologram and apply a mask of 0.25 nm$^{-1}$ on the side-band spot before calculating an inverse Fourier transform. \\
To separate the electrostatic and magnetic contributions to the phase shfit, two holograms were recorded before and after inverting the sample. Image calculations were then performed to aligned the two images. The phase images have been digitally flipped for accordance with the physical inversion of the sample. After data acquisition, an accurate correction of the drift, rotation and scaling between the two images has been performed using recently developped scripts.\\
Mean Inner Potentials have been calculated using the Doyle and Turner scattering amplitude corrected with the Ross and Stobbs equation (see chapter 12 of.\cite{volkl_e_introduction_1999}) We calculate : $V_{FePd_{L1o}} = 21.73$ V,  $V_{FePd_{dis.}} = 22.67$ V,  $V_{Pd} = 22.37$ V.\\
Micro-magnetical simulation has been carried out using the bulk FePd following parameters : Exchange constant $A = 6.9 10^{-12} \;$ J.m$^{-1}$, Uniaxial Anisotropy K=$1.03 \: 10^{6} \;$ J.m$^{-3}$, Saturated Magnetization $\mu_{0}$M$_{S}$ = $1.294 \: \: $Tesla . The cells are $ 0.781 \: nm \times 0.625 \: nm$ and infinite along the \emph{z} direction (considered as invariant).

\end{suppinfo}

\bibliography{Biblio,Bilbio2}

\end{document}